%% file: main.tex
\documentclass[manuscript]{acmart}

\usepackage{makecell}
\usepackage{multirow}
\usepackage{xcolor}

\AtBeginDocument{%
  }

\copyrightyear{2027}
\acmYear{2027}
\setcopyright{cc}
\setcctype{by}
\acmConference[IUI '27]{32nd International Conference on Intelligent User Interfaces}{February 8--11, 2027}{Helsinki, Finland}
\acmBooktitle{32nd International Conference on Intelligent User Interfaces (IUI '27), February 8--11, 2027, Helsinki, Finland}

\NewDocumentCommand{\fb}{O{white} O{black} m}{%
  \begingroup
  \setlength{\fboxsep}{0.3ex}%
  \raisebox{0ex}{%
    \fcolorbox{black}{#1}{%
      \rule{0pt}{0.7em}%
      \textcolor{#2}{\textbf{#3}}%
    }%
  }%
  \endgroup
}

\usepackage{titlesec}
\titlespacing*{\paragraph}
{0pt}      
{0.4em}    
{0.4em}    

\titlespacing*{\section}
{0pt}      
{0.4em}    
{0.4em}    

\begin{document}


\title{[MM/AI] Mental Models in Human-AI Interaction: Methods and Challenges in the Generative and Agentic AI Era (Workshop)}
\author{Téo Sanchez}
\orcid{0000-0001-7221-7020}
\affiliation{%
  \institution{Ludwig Maximilian University}
  \city{Munich}
  \country{Germany}
}
\email{teo.sanchez@lmu.de}

\author{Bhada Yun}
\orcid{0009-0008-5461-0003}
\affiliation{%
  \institution{ETH Z{\"u}rich}
  \city{Z{\"u}rich}
  \country{Switzerland}
}
\email{bhayun@ethz.ch}

\author{Prerna Ravi}
\orcid{0000-0002-4289-5610}
\affiliation{%
  \institution{CSAIL, Massachusetts Institute of Technology}
  \city{Cambridge, MA}
  \country{USA}
}
\email{prernar@mit.edu}

\author{Laura Schütz}
\orcid{0000-0002-5534-3903}
\affiliation{%
  \institution{Technical University of Munich}
  \city{Munich}
  \country{Germany}
}
\email{laura.schuetz@tum.de}

\author{Anna Neumann}
\orcid{0009-0000-9672-8087}
\affiliation{
    \institution{Research Center Trust, University Duisburg-Essen}
    \city{Duisburg}
    \country{Germany}
}
\email{}

\author{Robin Shing Moon Chan}
\orcid{0009-0007-1319-8334}
\affiliation{%
  \institution{ETH Z{\"u}rich}
  \city{Z{\"u}rich}
  \country{Switzerland}
}
\email{chanr@ethz.ch}

\author{April Yi Wang}
\orcid{0000-0001-8724-4662}
\email{april.wang@inf.ethz.ch}
\affiliation{%
  \institution{ETH Zurich}
  \city{Zurich}
  \country{Switzerland}}

\author{Qiaosi (Chelsea) Wang}
\email{qiaosiw at andrew.cmu.edu}
\affiliation{
\institution{HCII, Carnegie Mellon University}
  \city{Pittsburgh}
  \country{USA}
}
\author{Sumit Asthana}
\email{sumitasthana@microsoft.com}
\affiliation{
\institution{Microsoft}
  \city{Redmond}
  \country{USA}
}


\renewcommand{\shortauthors}{Sanchez, Yun et al.}

\begin{abstract}
\input{sections/0_abstract.tex}
\end{abstract}





\newcommand{\todo}[1]{\textcolor{red}{\textbf{TODO:} #1}}

\maketitle

{
}

\input{proposal.tex}

\end{document}

%% file: sections/0_abstract.tex
The mental model construct is widely used in HCI to refer to the knowledge structure people hold in order to reason about and interact with computing systems. Yet it is often operationalized intuitively: the construct is often used interchangeably with related concepts (e.g. folk theories, sensemaking) and methods of studying it (e.g. through elicitation) are many and diverse, with each method resting on distinct assumptions about what counts as a mental model. Generative and agentic AI systems may further complicate mental model formation and elicitation as such systems are opaque by design and increasingly act on users' behalf across files, applications, and on the web. Together, these challenges may hinder the commensurability of research on people’s mental models of AI systems. The MM/AI workshop calls for a critical reassessment of how we understand and study mental models in human-AI interaction research. It aims to foster theoretical and methodological exchange on mental models in human–AI interaction, identify open challenges, and develop directions for future research. We invite short papers on users’ or stakeholders’ mental models of AI systems, particularly contributions that reflect on the conceptual and methodological foundations of the construct. The half-day workshop combines lightning talks, hands-on elicitation exercises, and structured discussions on key questions concerning the future of the mental model for human-AI interaction research.

%% file: proposal.tex
\section{Description}

\paragraph{Background and motivation}

The use of the mental model construct in human-AI interaction research, including at IUI, has grown steadily since 2018~\cite{sanchez2026mental}. Borrowed from cognitive psychology~\cite{johnson-laird_mental_1983}, it refers to people's internal representation of a system, constructed through interaction and shaped by prior knowledge and expectations~\cite{carroll_mental_1988}. Mental models support people's ability to reason about and predict system behavior, and ultimately guide their actions. HCI and human factors research has shown that understanding a system's inner mechanisms can improve performance, learning, retention, skill transfer, and satisfaction~\cite{young_surrogates_2014,kieras_role_1984,payne_nature_1990,kulesza_tell_2012}. Conversely, knowing what mental models users actually hold is a means to improve the system's design. This has motivated work on mental model elicitation, i.e., capturing people's reasoning about a system in a form that can be examined by others~\cite{andrews_role_2023}.
However, the construct's relevance for our research community faces three important challenges.
First, studies elicit mental models in heterogeneous ways: \emph{researcher-led verbalizations} such as interviews and open-text questions~\cite{tullio_how_2007, brachman_building_2025, petridis2026compass}, \emph{participant-led verbalizations} such as think-aloud and teach-back~\cite{desolda_digital_2023, horstmann_alexa_2023}, \emph{structured judgments} such as comprehension scales and prediction tests~\cite{anderson_mental_2020, nourani_anchoring_2021, sungeelee_comparing_2024}, and \emph{artifact-based methods} such as free drawing and card sorting~\cite{kunkel_identifying_2021, li_how_2020, lee_hey_2022}. Each method rests on different assumptions about what constitutes a mental model, making it difficult to compare the results.
Second, the construct is often used interchangeably with adjacent concepts such as folk theories and sensemaking~\cite{jones_mental_2011}, blurring what exactly is being studied. 
Third, and most importantly, generative and agentic AI systems may challenge mental model formation and elicitation: these systems are opaque by design, adaptive, and increasingly act on users' behalf across files, applications, and the web. Misconceptions about their behavior can have concrete consequences. For instance, a user believing a coding agent cannot read files listed in a \textsc{.gitignore} may inadvertently send sensitive data to a third-party provider.
\textbf{These three challenges may ultimately hinder the legibility and commensurability of mental model research at IUI and beyond. We argue it is therefore timely to dedicate a space for theoretical and methodological exchange on the construct at IUI '27.}

\paragraph{Workshop goals}

This workshop aims to foster dialogue among researchers and practitioners about mental models in light of contemporary generative and agentic AI. It offers a venue for exchanging first-hand methodological experiences and perspectives on how people’s mental models of these systems can or should be studied. The workshop's objectives are threefold:\\
\noindent \textbf{Obj 1.$\quad$ Map elicitation practices.} Document and compare how participants conceptualize and study mental models of AI systems across different research contexts, through their paper submissions.\\
\noindent \textbf{Obj 2.$\quad$ Operationalize mental models for AI.} 
Surface methodological challenges and opportunities associated with studying mental models of generative and agentic AI through participants’ first-hand research experiences, as reflected in their submissions, and through hands-on elicitation exercises.\\
\noindent \textbf{Obj 3.$\quad$ Reflect on the future of the construct.} Advance discussion on how the mental model construct should be dynamically conceptualized and applied to remain relevant in the context of contemporary AI systems through structured group discussions.

\paragraph{Target audience}
The workshop targets researchers and practitioners, from academia or industry, ideally with first-hand experience with people's mental model of AI in the context of HCI.

\section{Previous history}
This is the inaugural edition of the workshop under the proposed organizing team.

\section{Organizers}

The organizing team combines first-hand research experience about mental models in human-AI interaction, experience in community-building activities, and a strong motivation to reopen and shape future research on these topics within the IUI community.

\begin{table*}[h!]
\centering
\footnotesize
\begin{tabular}{p{1.6cm} p{2.7cm} p{3.1cm} p{3.2cm} p{1.5cm} p{0.8cm}}
\toprule
\textbf{Name} & \textbf{Affiliation} & \textbf{Email} & \textbf{Web Page} & \textbf{Role} & \textbf{Attend.} \\
\midrule
Téo Sanchez & LMU Munich & \href{mailto:teo.sanchez@lmu.de}{teo.sanchez@lmu.de} & \url{https://teo-sanchez.github.io} & Main contact & Yes \\
\addlinespace
Bhada Yun & ETH Zürich & \href{mailto:bhayun@ethz.ch}{bhayun@ethz.ch} & \href{https://www.bhadayun.com/research}{https://bhadayun.com} & Co-organizer & Yes \\
\addlinespace
Prerna Ravi & MIT & \href{mailto:prernar@mit.edu}{prernar@mit.edu} & \url{https://prernaravi.com} & Co-organizer & Yes \\
\addlinespace
Laura Schütz & TUM & \href{mailto:laura.schuetz@tum.de}{laura.schuetz@tum.de} & \url{https://lauraschuetz.github.io/} & Co-organizer & Yes \\
\addlinespace
Anna Neumann & University Duisburg-Essen & \href{mailto:anna.neumann1@uni-due.de}{anna.neumann1@uni-due.de} & \url{https://annaneumann.carrd.co/} & Co-organizer & TBD \\
\addlinespace
Robin Chan & ETH Zürich & \href{mailto:chanr@ethz.ch}{chanr@ethz.ch} & \url{https://chanr0.github.io/} & Co-organizer & Yes \\
\addlinespace
April Yi Wang & ETH Zürich & \href{mailto:april.wang@inf.ethz.ch}{april.wang@inf.ethz.ch} & \url{https://peachlab.inf.ethz.ch/} & Co-organizer & Yes \\
\addlinespace
Qiaosi Wang & CMU HCII & \href{mailto:qiaosiw@andrew.cmu.edu}{qiaosiw@andrew.cmu.edu} & \url{https://www.qiaosiwang.me/} & Co-organizer & Yes \\
\addlinespace
Sumit Asthana & Microsoft & \href{mailto:sumitasthana@microsoft.com}{sumitasthana@microsoft.com} & \url{https://sumitasthana.xyz/} & Co-organizer & Yes \\
\bottomrule
\end{tabular}
\caption{Workshop organizers. Short biographies are provided in appendix~\ref{bios}.}
\label{tab:organizers}
\end{table*}

\section{Workshop Program Committee.}

Submissions will undergo double-blind peer review on \href{https://openreview.net/}{OpenReview}, supervised by a program committee.
Each submission will receive at least two reviews from a pool of reviewers comprising co-organizers and externally recruited researchers, with conflicts of interest managed through OpenReview. Program committee members will oversee review quality and making final recommendations. Besides quality, reviews will prioritize relevance to the workshop themes, methodological and theoretical insight, and potential to stimulate discussion.
Provisional members of the program committee are: \textbf{Prof. Dr. Simone Stumpf} (University of Glasgow), 
\textbf{Magdalena Wischnewski} (University of Duisburg-Essen), 
\textbf{Prof. Dr. April Yi Wang} (ETH Zürich), \textbf{Dr. Qiaosi (Chelsea) Wang} (Carnegie Mellon University), and \textbf{Dr. Sumit Asthana} (Microsoft research).\\

\section{Participants}
We expect 25–30 participants, including organizers, authors of accepted submissions, and attendees admitted through open registration. We welcome researchers and practitioners from HCI, AI, psychology, cognitive science, design, recruited through call for paper posted on a dedicated workshop website\footnote{\url{https://mentalmodelsofai.com/}}, and disseminated through ACM SIGCHI communication channels, the organizers’ professional networks, and social media. We invite short papers operationalizing mental models in human-AI interaction, particularly contributions discussing its conceptual and methodological foundations.

\section{Workshop activities}

The program is designed to progressively move from sharing participants' insight and identifying methodological challenges through paper presentations and hands-on exercises, toward reflecting on the future of the mental model construct in human-AI interaction through structured group discussion.

\noindent \textbf{\emph{0:00--0:15 --- Introduction (Dr. Téo Sanchez)}}: The workshop will begin with a brief introduction to its goals and themes, followed by a short genealogy of the mental model construct, tracing its origins in cognitive psychology and its adoption within HCI and human-AI interaction research.\\
\noindent \textbf{\emph{0:15--0:50 --- Selected lightning talks}}: Five submissions selected for their quality and diversity will be invited to present their work in a 5-minute lightning format, followed by 2 minutes for questions. The objective is to highlight the variety of ways mental models are conceptualized and operationalized across current research, while providing a common set of examples and points of reference for the subsequent activities.\\
\noindent \textbf{\emph{0:50--1:30 --- Peer elicitation exercise}}:
Participants will work in pairs where one will elicit the other's mental model of an AI system, focusing on one of two systems to facilitate comparison across groups: (1) a coding agent (e.g., Claude Code) or (2) a consumer-oriented text-to-image generator (e.g., DALL-E), two widely adopted and but contrasting forms of contemporary AI systems. Participants acting as elicitors will be free to choose among several elicitation approaches presented by the organizers. Method descriptions and supporting materials will be provided.\\
\noindent \textbf{\emph{1:30--2:00 --- Coffee break.}} If participants wish, the previous activity may continue during the coffee break.\\
\noindent \textbf{\emph{2:00--3:00 --- Breakout discussions}}: Participants will be assigned to breakout groups, each focusing on a different open question concerning the future of mental model research in the context of contemporary AI systems. Groups will draw on insights from the presentation, the elicitation exercise, and their own research experience. The goal is not necessarily to reach consensus, but to articulate competing perspectives and promising research directions. The questions under consideration are, and are not limited to: \textbf{Q1. Operationalization methods.} Can mental model elicitations be meaningfully compared within and across studies? What assumptions underlie different elicitation methods? \textbf{Q2. Contemporary AI systems.} How do contemporary AI systems challenge existing assumptions about mental models and their elicitation? What are (un)suitable class of methods for these systems? \textbf{Q3. Conceptual boundaries.} What makes the mental model construct distinctive from adjacent constructs such as folk theories, sensemaking, schemata, and anthropomorphism? \textbf{Q4. Dymanics.} How can researchers capture the dynamic nature of mental models given mental models continue to evolve through interaction with a system? \textbf{Q5. Research and design value.} What insights does mental model elicitation provide for the design of intelligent user interfaces? Could they be obtained through other approaches?\\
\textbf{\emph{3:00--3:30 --- Report-back and closing discussion}}: Each breakout group will briefly report their insights, points of disagreement, and the priorities they foresee for future research.

\section{Planned outcomes}

A public Zenodo archive will gather accepted papers, workshop materials, discussion summaries, and consented elicitation artifacts. The workshop website will stay online after the event and serve as a hub for accessing these materials.
We also plan to produce a short workshop report synthesizing key insights and future research directions identified during the discussions. The report will be deposited on Zenodo and HAL.

\section{Length}
Half-day




\bibliographystyle{ACM-Reference-Format}
\bibliography{sample-base, corpus}

\appendix

\section{Biographical and research backgrounds}
\label{bios}

\noindent\textbf{Téo Sanchez} is a MSCA Postdoctoral Fellow at the Munich Interactive Intelligence Initiative (MI3), LMU Munich. His HCI research rethinks users as proactive machine \emph{teachers} and investigates the psychological consequences of framing human-AI interaction as a teacher–student relationship. He recently conducted a systematic review of 88 empirical studies on mental models in human-AI interaction~\cite{sanchez2026mental}, which contributed to ground the framing of this workshop. His work has received awards at IUI 2022 (best paper), C\&C 2023 (honorable mention), as well as the AFIHM Best Ph.D. Dissertation Award in 2023.
\\
\noindent\textbf{Bhada Yun} is a PhD student at ETH Z\"urich studying Machine Intelligence and Visual \& Interactive Computing. His work develops phenomenological methods for studying human-AI interaction, eliciting how people subjectively perceive, make sense of, and relate to AI systems, with the aim of aligning AI to human wellbeing and agency. His research has been recognized with four CHI Honorable Mention Awards.
\\
\noindent\textbf{Prerna Ravi} is a PhD student at MIT's Computer Science and Artificial Intelligence Laboratory (CSAIL). Her research focuses on designing generative AI agents that augment team collaboration in education, creative practice, and collective decision-making. She has studied how users' mental models of conversational AI evolve over time, and how they both shape and are shaped by perceptions of the agent's trustworthiness, agency, and anthropomorphism. Her research has been recognized with the Best Paper Honorable Mention and Nomination awards at ACM CHI and Learning at Scale. Prerna was the main organizer for a tutorial workshop at the International Society of Learning Sciences (ISLS) in 2023. 
\\
\noindent\textbf{Laura Schütz} is a PhD candidate in Computer Science at Technical University of Munich. She also holds an MS in Design from Stanford. Her research focuses on modeling human perception and cognition for adaptive user interfaces. Laura is an incoming Postdoc Fellow at the ETH AI Center, where she will focus on learning user mental models from multimodal interaction traces during human-AI interaction. She has previously organized multiple interdisciplinary workshops, e.g., at ISMAR, Harvard Kennedy School, and Stanford Graduate School of Business.
\\
\noindent\textbf{Anna Neumann} is a PhD candidate in Computer Science at the Research Center for Trustworthy Data Science and Security. Her research focuses on transparency and governance requirements of technical systems to give stakeholders meaningful agency and recourse over systems, which has been recognized with a CHI Best Paper Award. She is part of the Compliant and Accountable Systems Group, which also focuses on perceptions of technological systems towards better governance of them. Anna has organized workshops, retreats, and mini-conferences for the RC Trust Graduate School which encompasses around 35 PhD students. 
\\
\noindent\textbf{Robin Shing Moon Chan} is a PhD candidate in Computer Science at ETH Zürich. His research focuses on understanding linguistic user behavior during LLM-assisted task solving and designing interactive algorithms and systems that effectively support such human–LM collaboration.
His work has been recently recognized with a CHI Best Paper Award. He has previously co-organized a tutorial workshop at ACL attended by 100+ NLP researchers.
\\
\noindent\textbf{April Yi Wang} is an Assistant Professor of Computer Science at ETH Zurich, where she leads the Programming, Education, and Computer Human Interaction Lab (PEACH) and is the chair of ACM SwissCHI local chapter. Through projects on AI tutors, classroom chatbots, learner-LLM problem decomposition, and agency in AI-mediated software engineering, she examines how educational tools can scaffold accurate mental model formation and repair rather than obscure it. 
She has served on the organizing committee of IEEE VL/HCC and regularly on the program committees of CHI, UIST, and Learning@Scale, as well as organized several workshops at CHI, IUI, and other venues.
\\
\noindent\textbf{Qiaosi (Chelsea) Wang} is a Carnegie Bosch Postdoctoral Fellow at Carnegie Mellon University, and an incoming assistant professor at UC Berkeley's School of Education. Her work focuses on theorizing, designing, and examining people's perceptions and mental models of conversational AI agents through the socio-cognitive lens of Mutual Theory of Mind across online learning, everyday life, and recently mental health contexts. Chelsea has served on the program committees of premier HCI conferences such as CHI, DIS, and CSCW. She's also the lead organizer of the Theory of Mind in Human-AI Interaction (ToMinHAI) workshop series at CHI and CUI.
\\
\noindent\textbf{Sumit Asthana} is a Senior Applied Scientist at Microsoft. His work focuses on advancing AI assistants and agentic systems, with a focus on human-AI interaction, decision support, evaluation frameworks, and synthetic data generation for AI-assisted experiences. His research spans human-centered AI, Bayesian modeling, AI-assisted decision-making, and the alignment of AI systems with human mental models. He has served on committees related to Human-centered AI in HCI and ML conferences such as CHI, CSCW, ICLR, NeurIPS.